\documentclass[preprint2]{aastex}
\input psfig.sty

\newcommand{\nh}{N_{\rm H}}
\newcommand{\sax}{{\it BeppoSAX~}}
\newcommand{\xte}{{\it RXTE~}}
\newcommand{\exosat}{{\it EXOSAT~}}
\newcommand{\rosathri}{{\it ROSAT/HRI~}}
\newcommand{\rosatpspc}{{\it ROSAT/PSPC~}}
\newcommand{\rosat}{{\it ROSAT~}}
\newcommand{\asca}{{\it ASCA~}}
\newcommand{\ginga}{{\it Ginga~}}
\newcommand{\ipc}{{\it Einstein/IPC~}}
\newcommand{\iue}{{\it IUE~}}

\def \nh {N${\rm _H}$}

\begin{document}

\title{Weak Reprocessed Features in the Broad Line Radio Galaxy 3C~382}

\author{Paola Grandi}
\affil{Istituto di Astrofisica Spaziale, CNR,
Area di Ricerca di Tor Vergata, Via Fosso del Cavaliere 100, I-00133 Roma, 
Italy}
\email{paola@ias.rm.cnr.it}
\author{Laura Maraschi} 
\affil{ Osservatorio Astronomico di Brera, Milano, Italy}
\email{maraschi@brera.mi.astro.it}
\author{C. Megan Urry}
\affil{Space Telescope Science Institute, Baltimore, USA}
\email{cmu@stsci.edu}
\author{Giorgio Matt}
\affil{Universita' degli Studi ``Roma tre'', Roma, Italy}
\email{matt@amaldi.fis.uniroma3.it }

\begin{abstract}
We present a detailed X-ray study of the Broad Line Radio Galaxy 3C~382,
observed with the \sax~satellite in a very bright state.
The continuum emission is well modeled with a power law that
steepens at high energies, with an e-folding energy 
of $\sim120$~keV.
At soft energies a clear excess of emission is detected, which
can not be explained solely by the extended thermal halo seen
in a \rosathri~image. 
A second, more intense soft X-ray component, possibly related to 
an accretion disk, is required by the data. 
Both a reflection component ($\Omega/2\pi=0.3$) and an 
iron line (EW$\sim 50$) are detected, at levels much
weaker than in Seyfert galaxies, suggesting a common origin.
Combining our measurements with results from the literature 
we find that the iron line has remained approximately constant over 
9~years while the continuum varied by a factor of 5. Thus
the fluorescent gas does not respond promptly to the variations of the X-ray 
primary source, suggesting that the reprocessing site
is located away, likely at parsec distances.
While the continuum shape indicates that X-rays derive from a thermal
Comptonization process, the weakness of other spectral features implies 
that either the upper layers of the optically thick accretion disk
are completely ionized or
the corona above the disk is outflowing with mildly relativistic velocity.
\end{abstract}


\section{Introduction}

Recent observations with \sax~(Grandi 2001, Zdziarski \& Grandi 2001) 
and \xte~(Eracleous, Sambruna \& Mushotzky 2000) have shown that the
circumnuclear environments in Broad-Line Radio Galaxies (BLRGs) 
are different from those of their radio-quiet counterparts, Seyfert~1 galaxies.
In particular, in radio-loud AGNs the reprocessed features can be
absent or weak, while in radio-quiet AGNs the iron lines and the 
reflection components are generally intense and always detected (Matt 2001).
These results have important implications for understanding
the physical distinction between radio-quiet and radio-loud AGNs,
that is why strong radio sources (lobes, jets, etc.) arise somehow only in
a small fraction  of  AGNs.

Broad Line Radio Galaxies share some of the properties of radio quiet
Seyfert galaxies, i.e. broad optical emission lines and an optical UV bump
suggesting the presence of a similar optically thick accretion disk in 
both cases. Yet the accretion disks may differ in the physical conditions 
in their inner regions: hence the diagnostic importance of X-ray spectra which
originate much closer to the central black hole.

The inner accretion disk in radio galaxies may be hot, optically
thin and  dominated by Advection (ADAF)
as proposed  early by  Rees et al. (1982) 
and studied  in detail more recently (Narayan 1997; Narayan, Mahadevan \& Quataert 1988).
The double-peaked broad lines observed in some BLRGs
can be explained if the lines are produced in a cold outer disk
surrounding a hot ion torus (Eracleous \& Halpern 1984).

X-ray observations of a narrow (although intense) iron line and a weak 
UV bump in 3C~390.3, the prototype of double-peaked radio galaxies, support 
the idea that cold accreting  material is not present in the vicinity of
 the black hole (Eracleous, Halpern \& Livio 1997; Grandi et al. 1999). 
A hot accretion flow might also explain the \sax~spectrum
of 3C~120, which shows a weak iron line and reflection component 
(Zdziarski \& Grandi 2001).

However the weakness of discrete features in the X-ray spectra of radio 
galaxies could also be due to dilution of a "normal" Seyfert continuum
by a pure power-law X-ray component possibly due to non thermal emission from the jets known to be present in these systems.

This hypothesis was not favored however
in the cases of 3C~390.3 and 3C~120 (Grandi et al. 1999, Grandi 2001, 
Zdziarski \& Grandi 2001). In 3C390.3 the iron line, although narrow is 
rather intense. In 3C120 the high energy cutoff observed by \sax~  
excludes a significant power law jet component.

Here we discuss in detail the \sax observation of the Broad Line Radio Galaxy
3C~382 (z=0.0579).
Preliminary results, together with an overall review of \sax~ spectra of BLRGs
were presented in Grandi (2000).
3C382 is a radio galaxy with lobe-dominated FRII morphology.
The inclination angle of its  radio jet is estimated by
Eracleous and Halpern (1988) to be  $i>15^\circ$, assuming 
the radio structure is not bigger than the largest double-lobed radio galaxies.
A relatively large inclination is supported by the low ratio of core 
to extended radio flux,
$R\equiv F_{core}/(F_{total}-F_{core}) = 0.07$ at 
6 cm (Rudnick, Jones \& Fiedler 1986), as compared for instance to $R=2.69$ 
for the well known OVV quasar 3C~279 (Morganti, Killeen \& Tadhunter 1993).
At optical-UV wavelengths, 3C~382 has broad (FWZI$>25,000$~km~sec$^{-1}$) 
 emission lines which are variable on time scales of months to years.
Its nuclear continuum is also strong and variable (Yee \& Oke 1981, 
Tadhunter, Perez \& Fosbury 1986). A recent HST WFPC2 image shows that
3C~382 is an elliptical galaxy strongly dominated by an unresolved 
nucleus (Martel et al. 1999).

The X-ray spectral properties of 3C~382 are not yet well established.

Previous satellites produced contradictory results on the soft emission.
Some authors found signature of warm absorber ( Nandra \& Pounds 1994, 
Reynolds 1997), others an excess of emission that was 
fitted with both a broken power law and  a thermal model (Urry et al. 1989,
Kaastra et al. 1991, Barr $\&$ Giommi 1992, Wo\'zniak et al. 1998, Sambruna et al. 2000).
Recently Prieto (2000) claimed that the soft excess could be entirely 
related to an extended thermal emission seen around the nucleus of 3C~382
in a \rosathri~ image.

At higher energies, the situation is also more confused.
Both \ginga~ and \asca~ and RXTE have detected 
an iron line, although with different profiles and 
intensities (Nandra \& Pounds 1994, Lawson \& Turner 1997, Reynolds 1998,
Wo\'zniak et al. 1998, Eracleous et al. 2000).
The \asca~iron line was extremely broad ($\sigma\sim 1.8$~keV) and more 
intense by about a factor 3 than that measured by \ginga.

The following sections describe the \sax observations, data analysis,
and interpretation. 

\section{\bf Observations and Data Reduction}

3C~382 was observed on 1998 September 20-22 
with the \sax Narrow Field Instruments (NFS): LECS (0.1-10~keV), 
MECS (1.5-10~keV), and PDS (15-300~keV).  
The LECS (Low Energy Concentrator Spectrometer) and MECS 
(medium Energy Concentrator Spectrometer) are imaging instruments and the PDS is a 
phoswich detector system with collimators pointing on and off the source.
At the time of the 3C~382 observations, only two units (MECS2 and MECS3)
of the MECS were working.
Table 1 shows the net exposure times and count rates of LECS MECS and PDS, 
respectively.

\begin{deluxetable}{lccccccc}
\tabletypesize{\small}
\tablecolumns{8}
\tablecaption{The observation log}
\tablehead{\colhead{Start time} &\colhead{End time}&\multicolumn{3}{c}{Exposure [s]}&\multicolumn{3}{c}{Count rate [s$^{-1}$]\tablenotemark{a}}\\
\colhead{} & \colhead{} &\colhead{LECS} &\colhead{MECS} &\colhead{PDS}
&\colhead{LECS} &\colhead{MECS}&\colhead{PDS}\\ }
\startdata
1998-09-20 17:01:18 &
1998-09-22 18:43:24 &
34572 &
85596 &
41613 &
$0.4863\pm 0.004$
& $0.7702\pm 0.003$ &
$0.84\pm 0.03$ \\
\enddata
\tablenotetext{a}{The energy ranges are 0.1--4 keV, 1.5--10 keV, 13--120 keV
for LECS, MECS, PDS, respectively. The count rates are background subtracted,
and their uncertainties are 1-$\sigma$. 
}
\label{t:log}
\end{deluxetable}

To test whether fast flux variations occurred,
we applied a $\chi^2$ test to the light curve for each instrument separately.
We set the  threshold for significant variation at a probability of
$\le10^{-3}$ that the count rate was constant,
independent of the temporal bins used
(we used bins of 1000, 5000 and 10000 seconds).
According to this criterion, no significant variations occurred during 
the \sax observation of 3C~382, even though
this source is known to be active on the time scale of a day 
(Barr \& Giommi 1982, Kaastra et al. 1991).
Note that flux variations of about $20\%$ as observed by \exosat could 
have been easily detected by the MECS instrument (see Figure 1).
Given the stationary state of the source, we summed the 0.1-120~keV 
spectrum over the entire observation.
\begin{figure}[ht]
\label{fig1}
\psfig{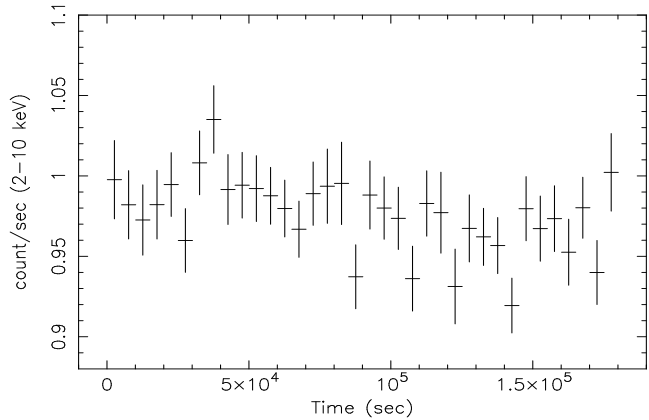}
\caption{The MECS light curve shows no significative evidence of fast 
variability. 
The bin size is 5000 sec and the length of the pointing about  2 
days.  The net exposure time is $\sim 86$ ksec: gaps in the data are due 
to Earth occultation and high background.} 
\end{figure}

Data from each instrument were reduced following
standard procedures (Fiore, Guainazzi \& Grandi 1999).
The LECS and MECS spectra were accumulated from circular regions of 4'.
A recent analysis by Prieto (2000) of the \rosathri images 
has demonstrated the presence of extended thermal emission around 3C~382.
As the FWHM is about 100 arcseconds, it is not resolved 
in the \sax images and therefore contributes to the LECS spectrum.
The \ipc~ data  revealed several
serendipitous point-like sources at distances of about 3' from 
the radio galaxy. Their contribution to
the \sax spectra should be  negligible since their \ipc count rates 
are about 30 times lower than those of 3C~382.
Even assuming a very flat power law $\Gamma= 1.5$ and Galactic \nh~
for the  serendipitous sources, their contribute to the PDS flux 
is less than 10\%. 
The LECS data were restricted to the 0.1-4 keV band because of  
well know calibration problems at higher energies (Grandi et al. 1997).

The PDS spectra were obtained combining the data from the 4 phoswich units.
Each pair of detectors is associated to one collimator.
The two collimators alternately point at the source and at the background.
We extracted PDS spectra using both fixed (FRT) and variable Rise Time (VRT)
thresholds to reject background (for details see 
Fiore, Guainazzi \& Grandi 1999). 
We checked that the second method did not alter the quality of 
the data or introduce a spurious deficit of high energy photons.
No substantial spectral difference between the FRT and VRT spectra
was found, so we used the Variable Rise Time spectrum
because of its better signal to noise ratio.
At the high energy end of the PDS band we excluded channels 
for which the signal-to-noise ratio was less than 3$\sigma$, limiting the
upper PDS energy to 120~keV.

Spectral data were then binned using the template files
distributed by the \sax~Scientific Data Center, in order to ensure
the applicability of the $\chi^2$ statistic and
an adequate sampling of the spectral resolution of each instrument.
LECS, MECS and PDS spectra were simultaneously fitted with XSPEC11,
with different relative normalization constants.
The MECS constant was fixed to the value 1 and the PDS one
was allowed to vary between 0.77-0.83. 
Since LECS and MECS data partially overlap, the LECS constant was not 
constrained, allowing the fit routine to find the best value.
As discussed by Fiore, Guainazzi and Grandi (1999), this procedure allows to 
correct the flux miscalibrations among the instruments.

In this paper, the reported uncertainties are at 90$\%$ confidence 
level for one parameter of interest ($\Delta\chi^2=2.71$), 
the fluxes are corrected for absorption and the corresponding
luminosities calculated assuming {\it H$_0=50$}~km~sec$^{-1}$~Mpc$^{-1}$
and {\it q$_0=0$}.

\section{\bf Results}

\subsection{Spectral Analysis}

The 3C~382 spectrum is complex and rich in features.
As shown in Figure 2, with respect to a power law 
absorbed by the Galactic column density,
\nh$=7.8\times10^{20}$ cm$^{-2}$ (Stark et al. 1992),
clear excesses are evident in the soft part of the spectrum ($<2 $~keV),
in the iron line region, and in the hard PDS band.
In order to better disentangle the different spectral components, we 
first analyzed the medium-hard X-ray spectrum (2-120~keV),
and then included the soft energy band.

A cutoff power law reprocessed by cold material (Magdziarz \& Zdziarski 1995; 
PEXRAV model in XSPEC) provides a good fit to the 2-120~keV  spectrum.
We fixed the smaller inclination angle allowed by the pexrav model ( 
$cosi= 0.95$) in agreement with the lower limit of the jet 
inclination $i>15^\circ$ inferred by Eracleous and Halpern (1998).
We also tested  larger inclination angle.
The fit is quite insensitive to this parameter for inclination 
angles $i<30^\circ$.

Note that a simple cutoff power law does not give a good fit to the data
($\chi^2=109.2$ for 81 degrees of freedom); 
$\chi^2$ is improved significantly by adding
a reflection component $R=0.5$ ($\chi^2=76.5$ for 80 d.o.f.).
An iron line, even if weak (EW=50~eV), is also required by the data;
parameterizing the line with a narrow gaussian profile ($\sigma_{Fe}=0$),
the fit improves significantly ($\chi^2=67.0$ for 78 d.o.f.,
$>99$\% significant according to the F-test).
Even if the intrinsic width of the iron line is let free to vary, 
the $\sigma$ value is small and consistent with zero (
$\sigma=0.09_{-0.09}^{+071}$ keV).

\begin{deluxetable}{lc}
\tabletypesize{\small}
\tablewidth{0pc}
\tablecolumns{2}
\tablecaption{\sax Fits\tablenotemark{a} to 0.1-120~keV Spectrum of 3C~382}
\tablehead{\colhead{} &\colhead{}}
\startdata
\multicolumn{2}{l}{\it Soft Component = Extended Thermal Emission Only}\\
&\\
\hline\\
$\Gamma_{hard}$ &1.89$^{+0.02}_{-0.03}$\\
E$_{\rm cutoff}$ [keV] &207$^{+190}_{-75}$\\
Reflection ($\Omega/2\pi$)&0.7$^{+0.3}_{-0.2}$\\
E$_{Fe}$ [keV]         &6.5$^{+0.2}_{-0.1}$\\
I$_{Fe}$\tablenotemark{b}&3.2$^{+1.7}_{-1.7}$\\
EW [eV]                &46$^{+26}_{-24}$\\
kT [keV]               &0.16$^{+0.04}_{-0.06}$\\
 $\chi^2$(dof)    &163(155)\\ 
&\\
Flux$^{\rm thermal~ component}_{\rm 0.1-2~keV}$ &1.0$^{+0.6}_{-0.4}$\tablenotemark{c}\\
&\\
Flux$^{\rm hard~ component}_{\rm 2-10~keV}$ &6.0$^{+0.2}_{-0.1}$\tablenotemark{c}\\
& \\
&\\
\hline\\
\multicolumn{2}{l}{\it Soft Component = Extended Thermal Emission + Power Law}\\
&\\
\hline\\
$\Gamma_{hard}$       &1.74$^{+0.11}_{-0.20}$\\
E$_{\rm cutoff}$ [keV] & 127$^{+99}_{-44}$\\
Reflection ($\Omega/2\pi$) &0.3$^{+0.3}_{-0.2}$\\
E$_{Fe}$ [keV]         & 6.5$^{+0.2}_{-0.1}$\\
I$_{Fe}^a$               &3.4$^{+1.7}_{-1.7}$\\
EW [eV]             & 48$^{+24}_{-25}$\\
kT [keV]& 1.1$^{+0.9}_{-1.7}$\\
$\Gamma_{soft}$&2.9$^{+0.6}_{-0.4}$\\  
$\chi^2$(dof)     &150(153) \\
&\\               
Flux$^{\rm thermal~ component }_{\rm 0.1-2.4~keV}$ &0.15$^{+10.50}_{-0.15}$\tablenotemark{c}\\
&\\
Flux$^{\rm soft~ power~ law~}_{\rm 0.1-2.4~keV}$ & 6.4$^{+11.4}_{-4.7}$\tablenotemark{c}\\ 
&\\ 
Flux$^{\rm hard~ component }_{\rm 2-10~keV}$ &6.0$^{+0.4}_{-4.0}$ \tablenotemark{c}\\
\enddata
\tablenotetext{a}
{\nh~ is fixed at the Galactic value, \nh=$7.8\times10^{20}$ cm$^{-2}$. 
The continuum is modeled with a cutoff power law reflected by cold material.
The line is a Gaussian with intrinsic width $\sigma_{\rm Fe}=0$.}
\tablenotetext{b}{Intensity of the iron line in units of $10^{-5}$ 
photons cm$^{-2}$ sec$^{-1}$.}
\tablenotetext{c}{Unabsorbed flux ($\times 10^{-11}$ erg cm$^{-2}$ sec$^{-1}$).}
\end{deluxetable}

With the limited statistics it is not possible to accurately 
describe the high energy steepening of the power law.
In fact, a broken power law (BEXRAV Model in XSPEC)
with $\Gamma_{\rm soft}=1.86$, $E_{\rm break}=23~$keV and 
$\Gamma_{\rm hard}=2.1$ gives an acceptable fit as well
($\chi^2$=65.5 for 77 d.o.f.).
Also in this case, the reprocessed features 
were both weak ($R=0.5$, EW=50~eV) and the intrinsic width of the line, 
when free to vary, still consistent with zero. 

Finally we included the soft part of the spectrum.
The \rosathri data show that
3C~382 is surrounded by an extended hot gas (kT=0.6 (+0.4,-0.1)~keV; 
Prieto 2000). 
A re-analysis shows that the \rosatpspc spectrum (0.1-2.4 keV) is well 
fitted by a nuclear power law 
($\Gamma=1.7$ and $F_{PSPC}\sim 2\times10^{-11}$ erg cm$^{-2}$ sec$^{-1}$) plus a bremsstrahlung component  which 
could account for all the soft excess reported by previous satellites.
However, given the uncertainties in the HRI
calibration the flux in the extended component 
is not well known and the determination made using the ROSAT PSPC 
is ambiguous since it could include a unsolved soft nuclear component.
($F_{HRI}\sim1.0\pm0.7\times10^{-11}$ erg cm$^{-2}$ sec$^{-1}$ and
$F_{PSPC}=2^{1.8}_{-0.7}\times10^{-11}$ erg cm$^{-2}$ sec$^{-1}$).

In the light of this, we added a thermal model to the PEXRAV model plus 
a narrow ($\sigma=0$) iron line.
This gives a good fit to
the entire 0.1-120~keV spectrum but residuals are still present below 
2~keV, indicating the presence of another (or more complex) soft component.
Note also that the fitted soft temperature is much lower than required
for the extended thermal gas (see Table 2).

\begin{figure}
\label{fig2}
\psfig{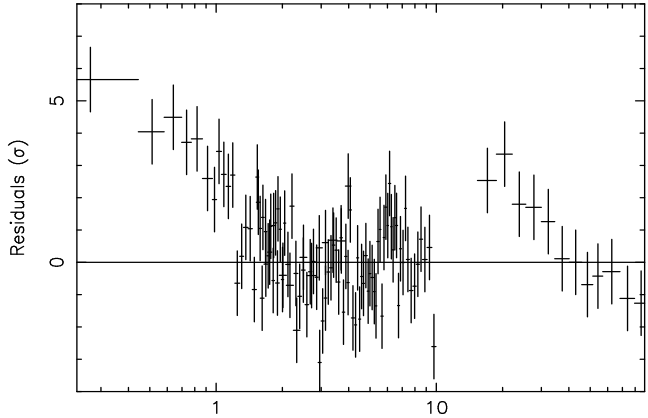}
\caption{ Residuals to a simple power-law model absorbed by Galactic \nh.
A strong soft excess is apparent below 1~keV, as are
an iron line and a hard excess above 10~keV.} 
\end{figure}

The addition of a soft power law ($\Gamma\sim 3$) 
significantly improves the $\chi^2$ (see Table~2; better
than 99.8$\%$ significance according to the F test).
Note that the thermal component is not required by the data any longer. 
Its flux is consistent with zero and the temperature cannot be constrained
very well (see Table 2). 
Even if the gas temperature is fixed to the ROSAT value 
(i.e. kT=0.6 keV), the flux of the extended component is still consistent
with zero.
During the \rosatpspc observation the nuclear component 
(modeled with a $\Gamma=1.7$ power law) was about a factor 2 lower that during our observation, 
so it is possible that the extended component is washed out
by the strong continuum and therefore only marginally detected by \sax.
(The extended component would not have varied on the time scale
of years.)
Other thermal models, such as a black body, can also
be used to fit the LECS excess instead of a simple power law.
However the poor spectral resolution of \sax below 1~keV makes this 
a useless exercise because the different models
are not statistically distinguishable.

We conclude that a hot extended plasma emission alone can not explain 
all the soft luminosity observed by \sax. 
Another softer and  stronger nuclear component is necessary. 

Once defined the best model for the entire  \sax spectrum 
(thermal emission plus and power law component plus cutoffed power law and reflection), we came back to the iron line 
study.  We allowed the intrinsic width ($\sigma$) of the line to vary.
Also in this case we found $\sigma$ consistent with zero, being 
estimated from the fits as 0.04 (-0.04,+0.48) keV. 

\subsection{Iron Line Historical Study}

One of the most striking results of the previous section is the weakness of 
the iron line and the reflection component, independent of what 
model was used to fit the continuum.

Since both the iron line and the reflection are weak, it makes sense to 
imagine the line arising in the same material that reflects the primary 
X-ray radiation. 
In order to understand the location of this reprocessing material, we 
investigated the iron line variability using data from the literature.

If the iron-emitting material is a standard cold disk with a 
hot  corona above it, 
a strong correlation between continuum and reprocessed features is expected.
Conversely, if either a non-thermal jet contaminates the X-ray continuum 
or the reprocessing gas is not near to the  primary X-ray source,
the continnum and line should not change together.

\begin{figure}
\label{fig3}
\psfig{figure=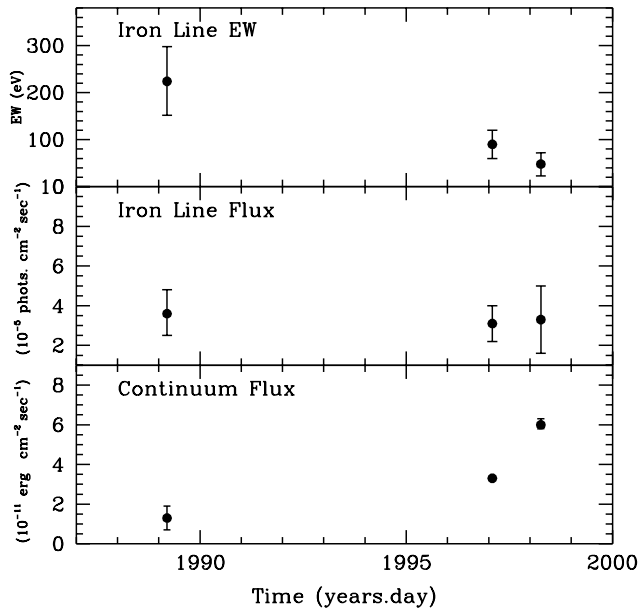,height=8.5cm,width=8.5cm}
\caption{2-10~keV continnum flux (lower panel), iron line flux (middle panel) 
and Equivalent Width (upper panel) of 3C~382 over 9 years of X-ray observation. Data are from 
\ginga (Wo\'zniak et al. 1998), \xte (Eracleous et al. 2000) 
and \sax (this paper). The reported error bars correspond to 90$\%$ 
confidence levels for one parameter of interest. 
The line flux appears independent  of the continnum level.}
\end{figure}

In Figure 3 the 2-10~keV continuum flux and
the iron line flux are plotted vs. the the date of observation.
The data are from \ginga (Wo\'zniak et al. 1998), \xte (Eracleous, 
Sambruna and Mushotzky 2000) and \sax (this paper).
We did not use the \asca data, in which the iron line was
broad and intense (Reynolds 1997).
As discussed by Zdziarski and Grandi (2001) for the 3C~120 case, 
the discrepancies between \sax/\xte and \asca are likely due to 
the \asca continuum being not well parameterized, either because of 
(i) the limited energy band covered by \asca, 
and consequently the inability to resolve all the spectral components,
or (2) calibration problems among the different instruments,
as reported by \asca team. It is known that the SIS CCDs suffer 
from serious degradation at energies below 1~keV and that the SIS and GIS 
results also diverge. These problems were already present, 
although not understood, since 1994
\footnote{\asca Calibration problems are reported on the web page
heasarc.gsfc.nasa.gov/docs/asca/wachout.html 
and discussed in the \asca-\sax intercalibration report available on 
the \sax webpage: www.sdc.asi.it/software.}.

Figure 3 shows clearly that, while the continuum flux changed by about a 
factor 5 over a 9-year period, the number of photons emitted in the
fluorescence line varied little (by less than a factor of 2).
This result has a very important implication: it shows that 
the cold matter does not respond promptly to variations of the primary 
continuum.

In principle, the increase in continuum flux could be due to
beamed jet radiation which dilutes the Fe emission of a 
stationary underlying (Seyfert-like) accretion disk.
As discussed in the next section, there are other (radio-optical-UV)
results which argue against the jet hypothesis in this source.
Therefore we interpret Fig.~2 as 
implying  a spatial separation between X-ray source 
and reprocessing matter. 

Note that similar results, based on repeated \sax observations, 
have been obtained for another radio galaxy, Centaurus A (Grandi et al. 1998).
In this source, the iron line appeared more intense when the nuclear 
source was at the lower intensity level,  supporting the idea 
that the reprocessed features are produced far away from the 
primary X-ray source.

Given the scarce sampling of the continuum and the iron line flux,
it is difficult to estimate the effective temporal lag (and therefore
the distance between reprocessing gas and X-ray source).
However the historical data support the idea that 
the reprocessing site is located at light years (parsecs) 
away from the X-ray primary source and coincides with a dense molecular torus. 

Ghisellini, Haardt and Matt (1994) studied the contribution of 
an obscuring torus seen on axis to the X-ray spectrum and 
showed that, for column density 
\nh$\ge10^{24}$ cm$^{-2}$, an important fraction of the hard X-ray radiation 
is reprocessed. The produced iron line is narrow and weak (80-100 eV) and the 
reflection component is 29-50$\%$ of the total flux at 30~keV.
Our data agree with these predictions within the uncertainties.

In addition the \sax data indicate that line is narrow.
For any tested model of the continuum, the best fit parameter 
of the intrinsic width is always  less than  90 eV and consistent with zero.
A narrow line is expected if the emitting gas is slowly rotating matter
at parsec distances.

\section{Discussion}

The \sax X-ray spectrum of 3C~382 has provided a wealth of interesting 
results. We find:
\begin{itemize}
\item{(1)} 
a soft component not associated with the extended thermal emission 
detected by \rosat;
\item{(2)} 
a medium-hard continnum well modeled with a cutoff power law ($E\sim120$~keV) 
or a broken power law (E$_{break}$ $\sim 30$~keV); 
\item{(3)}
a weak narrow iron line; and
\item{(4)}
a weak reflection component.
\end{itemize}
Both the iron line and reflection component are significantly 
weaker that those observed in Seyfert 1 galaxies. In addition, 
X-ray spectra of 3C~382 taken from the literature show that
the continuum and the iron line do not change together.

\subsection{The Continuum Emission}

In order to interpret these results, the first question 
is, how is the X-ray continuum emission produced? 
In the most popular scenarios for radio-loud AGNs, 
there are two possible origins for the X-ray emission:
\begin{itemize}
\item{(i)}
Comptonization of seed photons by hot rarefied gas.
The reprocessed region can be a patchy corona above a thin optically 
thick disk (Haardt \& Maraschi 1991, 1993; Poutanen \& Svensson 1996)
or a hot inner accretion flow and a cold outer thin disk (Shapiro, Lightman 
\& Eardley 1976; Narayan, Mahadevan \& Quataert 1998).
\item{(ii)}
Synchrotron and inverse-Compton radiation from relativistic electrons 
in a collimated outflowing plasma. 
The seed photons responsible for the inverse Compton component 
can be the synchrotron photons originating within the jet
(SSC; Maraschi, Ghisellini \& Celotti 1992, Bloom \& Marscher 1993) 
or photons external to the jet, produced by the accretion disk and/or the 
Broad Line Regions (EC; Sikora, Begelman \& Rees 1994).
\end{itemize}
It is likely that the former process is at work in radio-quiet AGNs, 
the latter in blazars.

Broad-Line Radio Galaxies are believed to be intermediate objects because they 
show  broad optical-UV  lines but also contain well collimated radio jets.
The question is then whether the observer sees 
X-ray photons from the jet or the accretion flow radiation or both.
In 3C~382, there are hints that there is no strong jet 
contamination at optical-UV and X-ray wavelengths.
Yee and Oke (1981) found an optical nuclear continuum which followed 
a power law ($\propto\nu^\alpha$) with alpha=+1.2
and -0.7 shortward and longward H$_\beta$, respectively. They 
suggested that the upturn at the blue end could be the UV bump from
the accretion disk, a feature commonly observed in Seyfert galaxies.
Moreover the blue continuum in
3C~382 is only  weakly polarized, whereas strong polarization
would be expected if the contribution of beamed synchrotron radiation 
were significant (Antonucci 1984, Cohen et al. 1999). 
HST observations of broad-line FRII galaxies also support the idea of 
relatively weak optical jet emission: while optical cores are found in 
a number of these galaxies, including 3C~382, 
their luminosities are poorly correlated with the radio 
luminosities (Chiaberge, Capetti \& Celotti 2000).
Compared to FRIs with similar radio luminosities, BLRGs appear much 
brighter in the optical, consistent with emission from an accretion
disk contributing significantly in these sources.

\begin{figure}
\label{fig4}
\psfig{figure=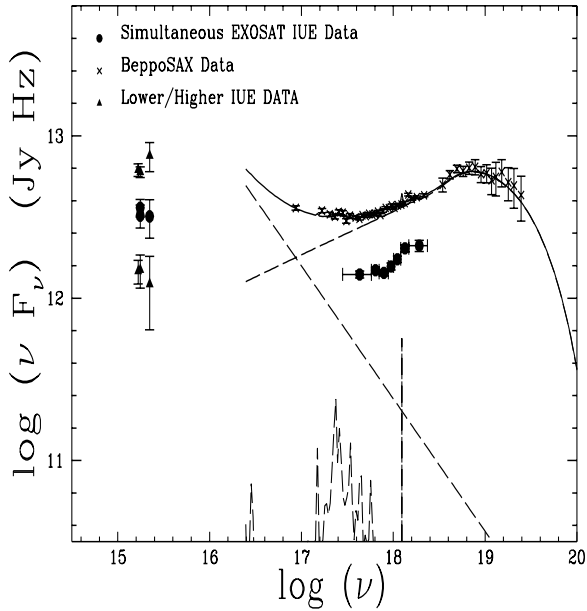,height=8.5cm,width=8.0cm,angle=-90}
\caption{The UV to X-ray spectral energy distribution of 3C~382.
The solid line represents the best fit to the \sax data (crosses),
as reported in Table~1, and the dashed lines are the
individual spectral components (e-folding power law plus 
reflection, iron line, soft power law, and weak thermal emission).
We suggest the UV bump and soft X-ray excess are produced by a 
cold accretion disk.
Filled circles represent an earlier, simultaneous \iue (Tadhunter et al. 1985)
and \exosat observation (Barr \& Giommi 1992), in which both the disk
emission and X-ray component are $\sim2$ times fainter.
Filled triangles represents the lowest and highest \iue fluxes
measured by Tadhunter et al. (1985). 
}
\end{figure}

Canosa et al. (1999) found a strong linear correlation between 
core radio luminosity (at 5~GHz) and unresolved \rosathri luminosity 
for a sample of B2 radio galaxies (mostly FRIs). 
(Similar results in Einstein data for a smaller and heterogeneous sample were 
found by Fabbiano et al. 1984)
Interestingly, 3C~382 is the only source in the B2 sample with
an X-ray luminosity about 100 times larger than the correlation
would predict. Canosa et al. (1999) argued that X-rays from 3C~382 must 
include strong X-ray emission from an unbeamed nuclear component, likely
related to the accretion disk.

Our \sax data, together with simultaneous \exosat and \iue data 
(collected on 1983 September 12; Barr \& Giommi 1992; 
Tadhunter et al. 1986), also support the idea of a strong 
disk contribution in 3C~382, as can be seen from the UV to X-ray 
spectral distribution in Figure~4.
The UV flux can change by more than a factor 7 on a time scale 
of years (Tadhunter et al. 1986).
Therefore, to represent the full range of UV variability, Figure~4 also
shows the lowest and highest fluxes measured by \iue.
The obvious UV/X-ray excess can be interpreted
as accretion disk emission plus a hard Comptonized tail extending into 
the soft X-ray band. 

Finally, we note that the spectral break at high energies 
observed in the \sax spectrum also argues against the blazar picture. 
According to unified schemes for radio-loud AGN (Urry \& Padovani 1995), 
BLRGs and intrinsically powerful blazars are the same objects 
seen at different viewing angles, in which case 
the jet output in the \sax energy range should 
be inverse Compton (Fossati et al. 1998, Ghisellini 1998). 
However, it is unlikely that the high energy cutoff in 3C~382 is 
a Compton peak, which in powerful blazars appears at MeV-GeV energies.

On the contrary, a power law which cuts off at high energies
is expected in Comptonization models, being 
related to the temperature of thermal electrons which 
scatter soft photons (Haardt \& Maraschi 1991, 1993; 
Poutanen \& Svensson 1997).
The e-folding energy for 3C~382 
is at a value typical of those for Seyfert 1s (Matt 2001).
We therefore conclude that the X-ray continuum in 3C~382 is produced
mainly by processes related to an accretion flow rather than non-thermal 
relativistic plasma in a jet.

\subsection{Weak Reprocessing Features}

The presence of a UV bump and soft X-ray excess 
indicates that the accretion material has to be cold 
and optically thick. 
When a cold slab is illuminated by a X-ray source, 
it produces an iron line and a reflected component
(Matt, Perola \& Piro 1991; George \& Fabian 1991),
which we should therefore expect to see in 3C~382.
If instead the bulk of the reprocessed radiation comes from 
much larger regions, as suggested by the lag between
continuum and iron line (Fig.~3), this means the cold accretion flow 
has somehow been inhibited from producing these characteristic 
spectral features.

One possible explanation is that the surface of the 
thin, optically thick disk (Shakura \& Sunyaev 1970) is highly ionized.
Nayakshin and Kallman (2000) have recently investigated how the
ionization of deeper and deeper layers of the accretion disk can affect 
the reprocessed features.
They showed that, if the X-rays are produced in magnetic 
flares at few disk scale heights, iron can be almost completely ionized 
providing the X-ray luminosity is a good fraction of the Eddington luminosity 
($L_{Edd}$). In that case, the skin of the disk acts 
as a perfect mirror and the reflected continuum is featureless.
Following their prescription, we calculated a lower limit for 
the accretion rate for 3C~382, $\dot{m}=L_{tot}/L_{Edd}>6\times10^{-3}$.
The total luminosity, $L_{tot}=L_{UV}+L_{X}\sim 5\times10^{45}$, was deduced 
from the simultaneous \iue and \exosat data reported in Fig.~4 assuming 
a high energy cutoff and a reflection strength as measured by \sax.
The Eddington luminosity was calculated assuming 
an upper limit for the black hole mass ($M<7\times10^9$ $M\odot$;
Tadhunter, Perez \& Fosbury 1986). 
In agreement with the magnetic flare geometry investigated 
by Nayakshin and Kallman (2000), the 
lower limit for the 3C~382 accretion rate is sufficient to obtain a 
completely ionized mirror-like disk.
Then the bulk of observed emission line can entirely come from a 
molecular torus.

Another possibility, which preserves the model of a cold disk, is that 
the hot corona has a mildly relativistic motion directed away from 
the disk (Beloborodov 1999a, 1999b).
A soft photon of energy $\epsilon_s$ passing through the hot corona acquires 
on average an energy A$\epsilon_s$. Beloborodov has calculated A as 
a function of the plasma bulk motion $\beta=v/c$ 
(see Fig.~1 in Beloborodov 1999a).
The spectral slope, $\Gamma$, depends on the amplification factor,
roughly as $\Gamma\sim 2.33(A-1)^{-0.1}$
(see also Fig.~2 in Beloborodov 1999b).
The spectral index $\Gamma=1.7-1.8$, as for 3C~382, 
implies a Comptonization factor $A\ge14$,
which in turn implies $\beta>0.4$ for hemisphere and slab 
geometries of the corona, i.e., a mildly relativistic flow.
The effect of the relativistic motion is to reduce the strength of 
the reflection. For $\beta\ge 0.4$ 
and small angle of view the radiation reflected ($R$) from the cold slab
becomes negligible ($R<0.1-0.2$).
In a similar way, mildly relativistic motions of the corona can influence
the iron line features. 
As shown by Reynolds and Fabian (1997), it is sufficient that a plasma flows
with $\beta\ge 0.4$ to reduce 
the equivalent width of the iron line by more than a factor 10.
Also in this case, if the disk production of the line is strongly inhibited, 
the main reprocessing  region can be a molecular torus. 

Note that an outflowing corona model could also account for the weak features of 
3C~120 (Zdziarski \& Grandi 2001). This hypothesis is particularly attractive 
for radio-loud AGNs. A hot plasma with a mildly relativistic velocity 
is not unlikely in sources  that  produce strong jets  in their nuclear regions.

\section{Conclusions}

The X-ray continuum of 3C~382 in a bright state  derived from a long
 \sax observation  is similar to that of  Seyfert galaxies,
i.e. a power law with index $\Gamma=1.74$,
with an e-folding cutoff at about 120~keV. 
At soft energies,  an excess of soft photons is detected  that can not be 
accounted for  by  an extended thermal component alone. The additional 
soft component required by the \sax data could be due to thermal emission from the innermost regions of a Shakura-Sunyaev disk.

Although a cold disk may be present, it must for some reason be
inhibited from producing features.
The iron line and the reflection are both detected, but 
they are weak and the iron line is unsolved by \sax.
This, plus the lack of correlation of the Fe line with the continuum flux,
suggests the cold reprocessing material is far from the primary 
X-ray source. It could be a dense molecular torus thought to surround the central source in many AGN ($N_H\ge 10^{24}$ cm$^{-2}$).
We discussed two possibilities that can explain the suppression  of features 
in the vicinity of the black hole: (i) the upper layers of the disk are 
completely ionized, or (ii) the patchy corona above the disk is not static
but flowing away with  mildly relativistic velocity.

\acknowledgements
We thank Ski Antonucci and Andrzej Zdziarski for valuable discussions. 
M. Guainazzi for analyzing the Einstein/IPC data.
This work was supported in part by NASA grant NAG5-9327.


\begin{references}
\reference{}
Barr P. \& Giommi P., 1992, \mnras, 255, 495
\reference{}
Beloborodov, A. M. 1999a, \apj, 510, L123
\reference{}
Beloborodov, A. M. 1999b, in ASP Con.\ Series 161, High Energy Processes in
Accreting Black Holes, ed.\ J. Poutanen \& R. Svensson (San Francisco: ASP),295
\reference{}
Blandford R. D., 1990 in Active Galactic Nuclei, Saas-Fee Advanvced Course, 
p. 264
\reference{}
Bloom, S. D., \& Marsher A.P., 1993, in Proceedings of CGRO AIP 280, Friedlander M., Gehrels N., Macomb D.J., eds, pag. 578
\reference{}
Canosa, C. M., Worral D. M., Hardcastle, M. J., Birkinshaw, M., 1999, \mnras, 310, 30
\reference{}
Chiaberge, M., Capetti A., Celotti A., 2000, \aa, 358, 104
\reference{}
Eracleous M., Halpern J.P., 1994, \apjs, 90, 1
\reference{}
Eracleous M., Halpern J.P., 1998, \apj, 550, 557
\reference{}
Eracleous M., Halpern J.P., Livio M., 1996 ApJ, 459, 89
\reference{}
Eracleous, M., Sambruna, R., \& Mushotzky, R. F. 2000, \apj, 537, 654 
\reference{}
Fabbiano G., Miller, l., Trinchieri, G., Longair, M., Elvis, M. 1984, \apj, 277, 115
\reference{}
Fiore, F., Guainazzi, M., \& Grandi, P. 1999, Cookbook for BeppoSAX NFI
Spectral Analysis\\
(www.sdc.asi.it/software/cookbook) (F99)
\reference{}
Grandi P., Tagliaferri, G., Giommi P., Barr, P., \& Palumbo, G. G. C.,
1992, \apjs, 82, 116
\reference{}
Grandi, P., Guainazzi, M., Mineo, T., Parmar, A.N., Fiore, F.,
Matteuzzi, A., Nicastro, F., Perola, G.C., Piro, L., Cappi, M., Cusumano, G.,
Frontera, S., Giarrusso, F., Palazzi, E., Piraino, S., 1997, A\&A, 325, L17
\reference{}
Grandi, P., Guainazzi. M., Haardt, F., Massaro, E., Matt, G., Piro, L.,
Urry C. M., 1999), A\&A, 343, 33
\reference{}
Grandi, P., Guainazzi. M., Haardt, F., Massaro, E., Matt, G., Piro, L.,
Urry C. M., 1999), A\&A, 343, 33
\reference{}
Grandi P., 2001, proceedings of {\it X-ray Astronomy 1999: Stellar Endpoints, AGN and the Diffuse Background}, eds. Palumbo G., Malaguti P., White, N., in press (astro-ph/0005577)
\reference{}
Haardt,~F., Maraschi,~L., 1991, \apj, 380, L51
\reference{}
Haardt~F., Maraschi~L., 1993, \apj, 413, 507
\reference{}
Kaastra J.S., Kunieda H., Awaki H., 1991,  A\&A, 242, 27
\reference{}
Magdziarz \& Zdziarski A. A, 1995 MNRAS, 273, 837
\reference{}
Maraschi L., Ghisellini, G., Celotti A.,1992, \apj, 397, L5
\reference{}
Martel A. R., Baum, S. A., Sparks W. B., Wyckoff, E., Biretta J. A.,
Golombek, D., Macchetto F. D., de Koff, S., McCarthy, P. J., Miley, G. K., 1999, \apjs, 122, 81
\reference{}
Matt G., 2001, proceedings of {\it X-ray Astronomy 1999: Stellar Endpoints, AGN and the Diffuse Background}, eds. Palumbo G., Malaguti P., 
White, N., in press
\reference{}
Meier D. L., 2000, New Astronomy Reviews, in press (astro-ph/9908283)
 \reference{} 
Morganti, R., Killeen N.E.B., Tadhunter, C.N., 1993, \mnras, 263, 1023 
\reference{}
Nandra K., Pounds K.A., 1994, MNRAS, 268, 405
 \reference{}
Nayakshin S., \& Kallman T. R., 2000, \apj, in press NK
Poutanen, J., \& Svensson, R. 1996, \apj, 470, 249
\reference{}
Prieto, M. A. 2000, \mnras, 316, 442
\reference{}
Rees M.J., Begelman M.C., Blandford R.D., Phinney E.S., 1982, Nature, 295, 17
\reference{}
Reynolds C., 1997, \apj, 286, 513
\reference{}
Rudnick L., Jones, T.W., Fiedler, R., 1986, \aj, 91, 1011
\reference{}
Shakura,~N.~I., Sunyaev, R.~A.~, 1973, A\&A  , 24, 337
\reference{}
Shapiro,~S.~L., Lightman~A.~P., Eardley~D.~M., 1976, \apj, 204, 
187
\reference{}
Sikora M., Begelman, M.C. \& Rees, M.J., 1994, \apj, 421, 153
\reference{}
Stark, A.A., Gammie, C.F., Wilson, R.W.,m Bally, J., Linke, R.A., 
	Heiles, C., Hurwitz, M., 1992, \apjs, 79, 77
\reference{}
Tadhunter,~C.~N., et al., 1986, MNRAS, 219, 55
\reference{}
Urry, C. M., \& Padovani, P. 1995, PASP, 107, 803
\reference{}
Urry C. M., et al. 1989, proceedings of the 23rd ESLAB Symposium, (ESA SP-296),
p. 789
\reference{}
Wilson, A. S., \& Colbert, E. J. M. 1995, \apj, 438, 62
\reference{}
Wo\'zniak, P. R., Zdziarski, A. A., Smith, D., Madejski, G. M., \& Johnson,
W. N. 1998, \mnras, 299, 449 (W98)
\reference{}
Zdziarski A.A. \& Grandi P., 2001, \apj,  in press.
\end{references}
\end{document}